\documentclass[aps,prx,superscriptaddress,twocolumn,longbibliography]{revtex4-1}

\usepackage{ucs}
\usepackage[latin1]{inputenc}
\usepackage[english]{babel}
\usepackage[usenames, dvipsnames]{xcolor}
\usepackage{amsmath, 
			amsfonts,
            booktabs,
            microtype,
            mathtools, 
            graphicx,
            color,
            hyperref,
            nicefrac,
            dsfont,
            url, 
            placeins,
            braket,
			appendix} 

\begin{document}

\title{Uncovering the non-equilibrium phase structure of an open quantum spin system}
\author{S. Helmrich}\affiliation{Physikalisches Institut, Universit\"at Heidelberg, Im Neuenheimer Feld 226, 69120 Heidelberg, Germany.}
\author{A. Arias}\affiliation{Physikalisches Institut, Universit\"at Heidelberg, Im Neuenheimer Feld 226, 69120 Heidelberg, Germany.}
\author{S. Whitlock}\email[e-mail: ]{whitlock@ipcms.unistra.fr}\affiliation{Physikalisches Institut, Universit\"at Heidelberg, Im Neuenheimer Feld 226, 69120 Heidelberg, Germany.}\affiliation{IPCMS (UMR 7504) and ISIS (UMR 7006), University of Strasbourg and CNRS, 67000 Strasbourg, France}
\date{\today}
	
\begin{abstract}
We experimentally and theoretically investigate the non-equilibrium phase structure of a well-controlled driven-disspative quantum spin system governed by the interplay of coherent driving, spontaneous decay and long-range spin-spin interactions. We discover that the rate of population loss provides a convenient macroscopic observable that exhibits power-law scaling with the driving strength over several orders of magnitude. The measured scaling exponents reflect the underlying non-equilibrium phase structure of the many-body system, which includes dissipation-dominated, paramagnetic and critical regimes as well as an instability which drives the system towards states with high excitation density. This opens up a new means to study and classify quantum systems out of equilibrium and extends the domain where scale-invariant behavior may be found in nature. 

\end{abstract}
	
\maketitle

\section{Introduction}

Statistical mechanics provides a powerful framework for understanding and classifying states of matter close to thermal equilibrium - a seminal example being the transition between paramagnetic and ferromagnetic phases of Ising magnets and the liquid-gas transition in fluids~\cite{Lee1952}. Close to their respective transition points, critical fluctuations with diverging correlations dominate, giving rise to remarkably simple scaling laws for macroscopic observables~\cite{sachdev2011}. As it happens, in the case of the Ising transition and the liquid-gas transition, these scaling laws involve just a few common exponents, indicating that both systems infact belong to the same universality class. 

Comparatively little is known about many-body systems in out-of-equilibrium scenarios~\cite{Hinrichsen2000}, especially open quantum systems, governed by a competition between quantum coherent evolution and dissipation. This is becoming especially relevant with the emergence of a new generation of experiments that are genuinely non-equilibrium in nature, such as crystals of laser cooled ions~\cite{Barreiro2011,Schindler2013,Bohnet2016}, semiconductor exciton-polariton condensates~\cite{Kasprzak2006}, ensembles of nitrogen-vacancy centers in diamond~\cite{Choi2017}, superconducting circuits~\cite{Houck2012,Eichler2015}, ultracold atomic gases in optical cavities~\cite{Brennecke2007,Colombe2007,Lin2011,Landig2016} and laser driven ensembles of Rydberg atoms~\cite{Carr2013,Schempp2014,Malossi2014,Valado2016,Melo2016,Weller16,Goldschmidt2016,Letscher2017}. An important feature of these systems is the interplay between coherent driving, dissipation (e.g., due to spontaneous decay) and interactions between the particles, that can give rise to fundamentally new states that are quite distinct from equilibrium matter. Classifying these new states of matter poses a significant challenge to state-of-the-art many-body theory and experiments, in part because theoretical methods capable of dealing with open many-body systems are less developed and because it is difficult to devise observables capable of distinguishing the vastly different types of behavior they can exhibit.

\begin{figure}[!t]
	\centering 
	\includegraphics[width = 0.9\columnwidth]{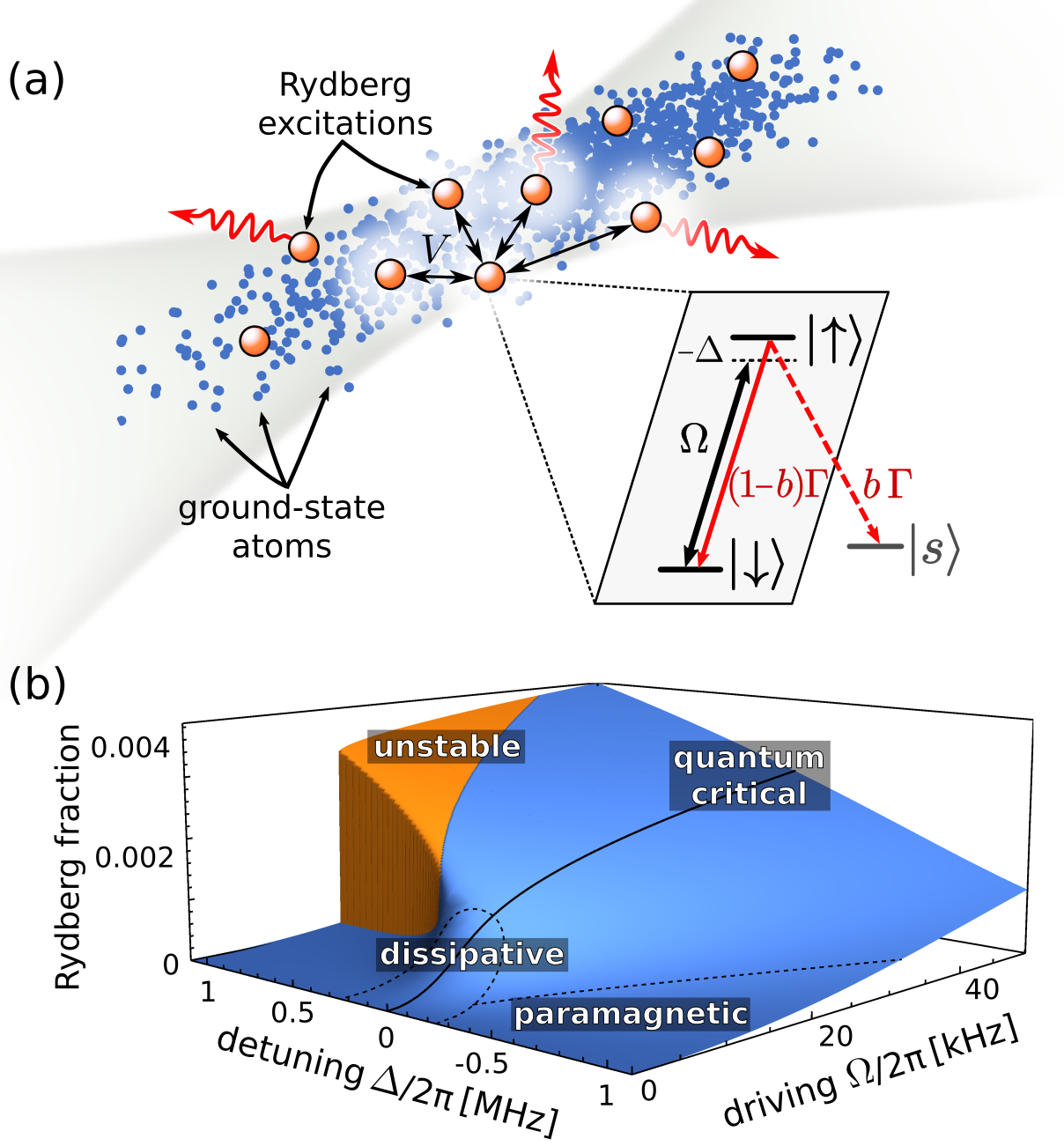}
	\caption{Prototypical open quantum spin system based on laser driven Rydberg atoms in an optical trap. (a) Geometry of the atomic gas (blue dots) with Rydberg excited atoms depicted as orange spheres. The ground and Rydberg states of each atom form a pseudospin-1/2  where the laser coupling ($\Omega$) and detuning ($\Delta$) play the role of transverse and longitudinal fields. Ising-like interactions arise from the repulsive van der Waals interactions between Rydberg states while dissipation enters via spontaneous emission of the excited state, which returns population either to the $\ket\downarrow$ state or out of the system to auxilliary shelving states represented by $\ket{s}$ (with branching factor $b$). (b) Non-equilibrium phase diagram obtained from mean-field theory showing the steady state fraction of Rydberg excitations $m$ (assuming $b=0$ and otherwise similar parameters to the experiment). The competition between driving, dissipation and interactions gives rise to a rich phase structure, including paramagnetic, dissipation dominated, critical and unstable regimes.} 
	\label{fig:1_regimes}
\end{figure}

Here we experimentally investigate the long-time dynamics of a widely tunable open quantum spin system (Fig.~\ref{fig:1_regimes}), over a wide range of driving field parameters. We show that the overall rate of population loss due to the decay of excited states provides a convenient macroscopic observable for the many-body state which can be measured with a dynamic range covering several orders of magnitude. We discover that this observable exhibits approximate power law scaling over a wide parameter range, with exponents that we associate to qualitatively different regimes. Our observations are in good agreement with theoretical modelling of the open-system dynamics based on coupled rate-equation simulations that include the effects of strong and long-range interactions between Rydberg excited atoms. Combining theory and experiment, we map out the non-equilibrium phase diagram of this system which exhibits four different regimes including dissipation-dominated and paramagnetic regimes as well as two distinct many-body regimes which arise through the competition between interactions and driving.

\section{Open quantum spin system}

Our system consists of a gas of ultracold atoms driven by a laser field to create a small fraction of short-lived Rydberg excitations [Fig.~\ref{fig:1_regimes}(a)]. A convenient way to describe the system is in terms of a quantum master equation in Lindblad form for the many-body density matrix $\rho$ (in units where $\hbar=1$):
\begin{equation}\label{eq:lindblad}
\partial_t\rho=\!- i[\mathcal H,\rho]+ \sum_j\!\left [L_j\rho L_j^{\dagger}\!-\!\frac{1}{2}\left(L_j^{\dagger} L_j \rho\!+\!\rho L_j^{\dagger} L_j \right)\right].
\end{equation} 
In this equation, the Hamiltonian $\mathcal H$ accounts for the coherent part of the dynamics, while the second term accounts for dissipative processes described by the set of quantum jump operators $L_j$. In the following we assume that the laser couples the ground and Rydberg states, while dissipation is primarily due to spontaneous decay of the Rydberg excitations either back to the original ground state or to auxiliary uncoupled states (denoted $\ket{s}$) that do not participate in any subsequent dynamics. Identifying the ground and Rydberg states as the spin-down and spin-up states of a pseudo-spin 1/2 and using the rotating wave approximation, we can express the Hamiltonian as
\begin{equation}\label{eq:hamiltonian}
\mathcal H = \frac{\Omega}{2}\sum_j \sigma_x^j - \frac{\Delta}{2} \sum_j \sigma_z^j + \frac{1}{2}\sum_{j,k\neq j} V_{jk} n^j n^k,
\end{equation}
where $\sigma_x^j$, $\sigma_z^j$ are Pauli spin matrices, $n^j = (\sigma^j_z+1)/2$ and the indices $j,k$ refer to each spin in the ensemble. In our experiment antiferromagnetic spin-spin interactions originate from the repulsive van der Waals interactions between Rydberg excitations. These interactions fall off as a power law $V_{jk}=C_6/|\vec r_j-\vec r_k|^6$ but due to the extremely large $C_6$ coefficients of Rydberg states they can extend far beyond nearest-neighbors. This has the consequence that a single excitation can suppress the subsequent excitation of hundreds of nearby spins within a characteristic volume called the Rydberg blockade volume. A convenient parameter which characterises the van der Waals interactions between neighboring Rydberg states is $J = C_6{\rho_0}^2$ where $\rho_0$ is the peak atomic density. The atom-light coupling strength $\Omega$ and the detuning from the atomic transition $\Delta$ correspond to transverse and longitudinal fields respectively, which can be tuned over a wide range via the Rydberg excitation laser. The quantum jump operators describing spontaneous decay are $L_j=\sqrt{\Gamma}\sigma_-^j$, with the spontaneous decay rate $\Gamma$. Additionally, single spin dephasing, e.g. to account for laser phase noise, can be included via additional jump operators of the form $L_j^\mathrm{de}=\sqrt{\gamma_{\rm{de}}}n^j$.

Neglecting dissipation for a moment, this system closely resembles the quantum Ising model in transverse and longitudinal fields~\cite{Weimer2008,Schauss2015,Labuhn2016}. Thus, in the limit $\Omega\rightarrow 0$ the resulting equilibrium ground state phase diagram includes a paramagnetic phase (for $\Delta <0$) and a hierarchy of crystalline phases (for $\Delta >0$) with varying excitation densities~\cite{Schauss2015}. Increasing the coherent laser coupling $\Omega>0$ introduces quantum fluctuations which lead to the appearance of an experimentally accessible quantum critical region~\cite{Heidemann2007,Loew2009}. However, the inclusion of spontaneous decay of the Rydberg states breaks the detailed-balance condition of equilibrium physics, which can have dramatic effects on the corresponding phase structure.

To help navigate the non-equilibrium phase structure, in Fig.~\ref{fig:1_regimes}(b) we present mean-field results for the steady state Rydberg fraction $m=\langle n\rangle$, which plays the role of the magnetization. For this we assumed a product state ansatz and a homogeneous system but we explicitly include a hardcore constraint for the two-point correlations  which captures the Rydberg blockade effect (further outlined in Appendix~\ref{app:mean_field}). Except for the orange region in Fig.~\ref{fig:1_regimes}(b), the system has a unique steady state corresponding to a small and smoothly varying fraction of Rydberg excitations. From further inspection of the mean-field solution we can identify four different regimes:

\noindent\emph{Paramagnetic regime} ($|\Delta|\gg \Omega,\Gamma$): For detunings far above or below resonance and for weak driving and dissipation, each spin aligns with the external field according to the relative strength of $\Omega$ and $\Delta$. For large detunings the Ising interaction term can be considered a small perturbation yielding a paramagnetic state with magnetization scaling as $m=\Omega^2/4\Delta^2$. This is equivalent to each atom being in the weakly-dressed state $\ket{\psi}\approx \ket{\downarrow}+\beta\ket{\uparrow}$ with $\beta=\Omega/(2\Delta) \ll 1$. 

\noindent\emph{Dissipation-dominated regime} ($\Delta\approx 0;\Gamma\gg\Omega$): Driving the system close to resonance, if the single-spin spontaneous decay (or dephasing) rate is large compared to the driving Rabi frequency, then this results in a continuous projective measurement of each spin in the $\sigma_z$ basis. Consequently, the system will evolve to a classical spin configuration comprised of a small but fluctuating number of spin-up excitations. In this limit the steady state Rydberg fraction scales as $m=\Omega^2/\Gamma^2$. 

\noindent\emph{Critical regime} ($\Delta\approx0; \Omega, J\gtrsim \Gamma$):
Also close to resonance, as the driving field strength is increased, the system undergoes a crossover from the dissipation-dominated regime to a high-density liquid-like state. This coincides with a change in the $\Omega$ dependence of the magnetization that originates from the critical regime associated to the quantum critical point of the equilibrium Ising-like model at $\Delta=\Omega=0$. Mean-field theory including hard-core correlations, predicts a scaling law for the magnetization $\propto \Omega^{\alpha}$, where $\alpha_\mathrm{MF}=2/5$ is one of the universal critical exponents of the model.

\noindent\emph{Unstable regime} ($\Delta\sim J, \Omega\gg\Gamma$):
The competition between dissipation, driving and interactions is perhaps most striking for intermediate detunings above resonance (i.e. the orange region of Fig.~\ref{fig:1_regimes}(b) for $\Delta>0$), where mean-field theory predicts a discontinuity in the magnetization and the appearance of a bistable region associated with a transition to an ordered phase. It is debated whether such bistabilities due to Rydberg-Rydberg interactions can be observed in experiments~\cite{Carr2013,Melo2016,Weller16,Letscher2017}, while theoretical studies of similar models taking into account beyond-mean-field corrections suggest that the bistable phase may be replaced by a first order transition and a tricritical point~\cite{Weimer2015,Overbeck2017}.

\begin{figure}[!t]
	\centering
	\includegraphics[width = 0.9\columnwidth]{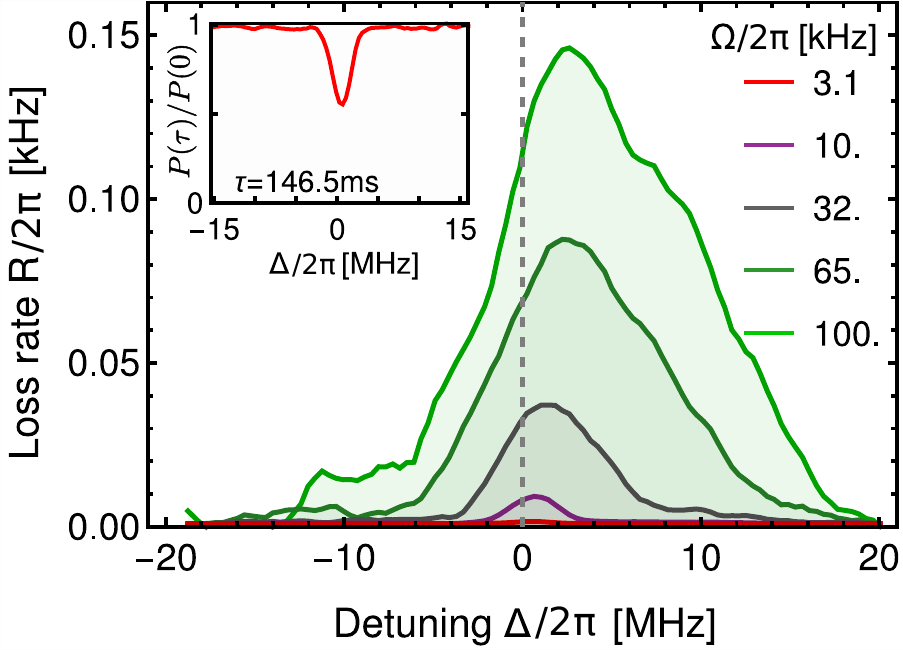}
	\caption{Measured loss rates as a function of detuning $\Delta$ for various driving field strengths $\Omega$. As $\Omega$ is increased the spectra become asymmetric and broaden due to strong and repulsive Rydberg-Rydberg interactions. Inset: Fraction of atoms $P(\tau)/P(0)$ remaining in $\ket{\downarrow}$ for the smallest driving field strength $\Omega/2\pi=3.1\,\rm{kHz}$.}
	\label{fig:2_spectra}
\end{figure}

\section{Probing the open system dynamics}

\begin{figure}[!t]
	\centering
	\includegraphics[width = 0.9\columnwidth]{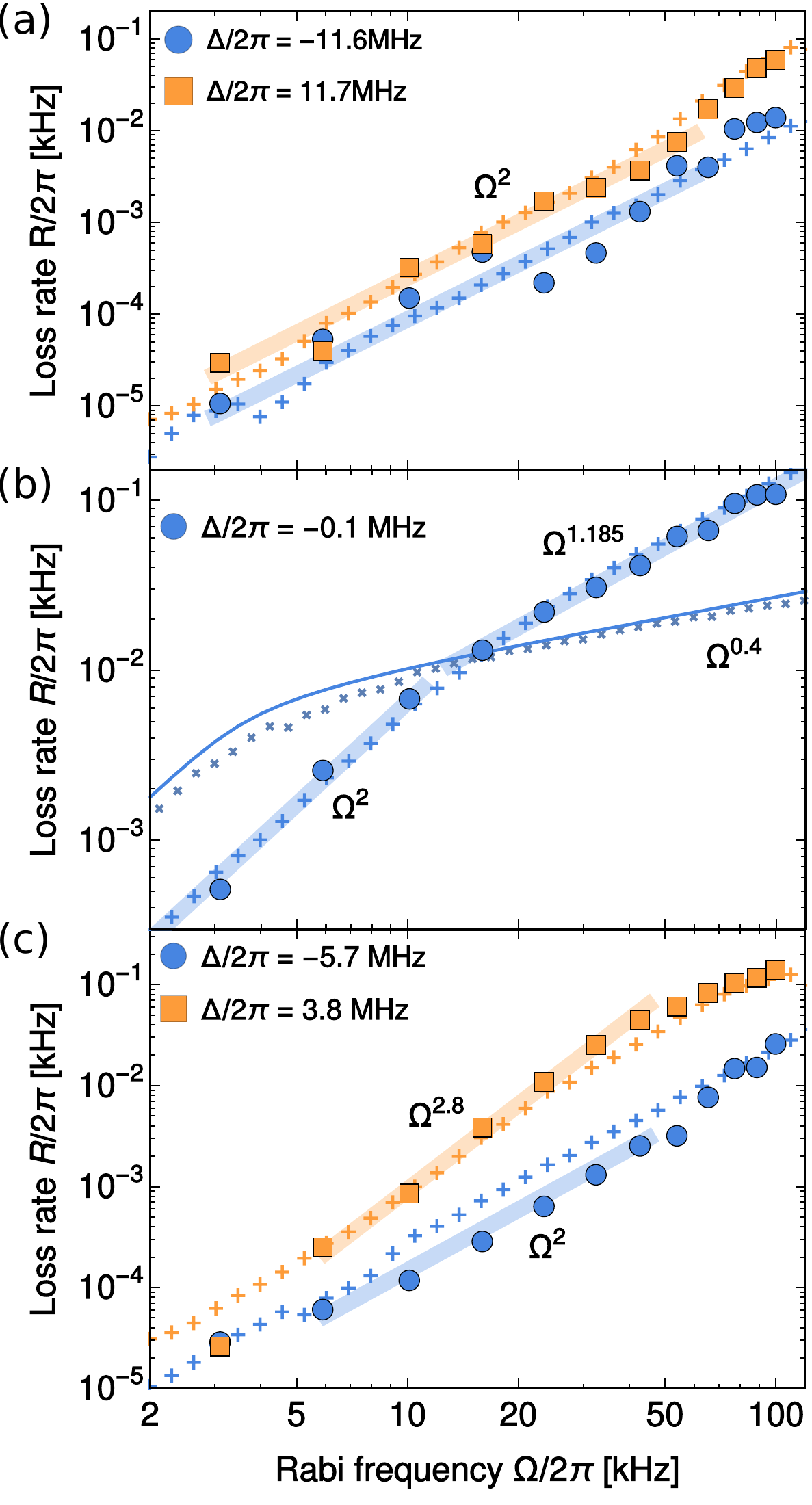}
	\caption{Power law scaling of the loss rate $R$ as a function of driving strength with different exponents that depend on the different regimes of the driven system. The thick shaded lines show the power-law scaling with exponents used to distinguish the different regimes discussed in the text. Rate equation simulation results are shown with $+$ symbols. In panel (b) we additionally show homogeneous results for the rate equation simulations (crosses) and for mean-field theory (solid line). (a) Far from resonance the loss rate exhibits paramagnetic scaling with $\alpha=2$. (b) Close to resonance the loss rate exhibits two different scaling regimes with $\alpha=2$ (dissipation-dominated) and $\alpha=1.185$ (critical). (c) For intermediate detunings we observe $\alpha>2$ attributed to facilitated excitation that tends to drive the system into the critical regime $\alpha <2$ for large $\Omega$.
	}
	\label{fig:3_scaling}
\end{figure}

To experimentally explore this rich non-equilibrium phase structure we perform experiments on a gas of $7.5\times 10^4$ $^{39}$K atoms initially prepared in the $\ket{\downarrow}=\ket{4s_{1/2},F=2,m_F=2}$ state and randomly distributed in a cigar shaped optical dipole trap. The peak atomic density and temperature are $\rho_0=5\times10^{11}\,\rm{cm}^{-3}$ and $T=19.4\,\rm{\mu K}$ respectively. All atoms in the sample are then driven from $\ket{\downarrow}$ to $\ket{\uparrow}=\ket{66s}$ by a two-photon laser excitation with large detuning from the intermediate state such that it can be mostly neglected (for details see Appendix~\ref{app:experimental_methods}). We express the laser parameters in terms of the effective two-photon Rabi frequency $\Omega/2\pi$, which we vary from $3\,\rm{kHz}$ to $100\,\rm{kHz}$, and the detuning $\Delta/2\pi$ which is varied over the range $\pm 20\,\rm{MHz}$ (Fig.~\ref{fig:2_spectra}). The strength of the nearest neighbour van der Waals interactions of Rydberg pair states is $J/2\pi = 65.4\,\rm{GHz}$, greatly exceeding all other energy scales such as those associated to single particle driving and decoherence, leading to strong blockade effects. The total effective decay rate of the excited states, including spontaneous emission and black body decay from the Rydberg and short-lived intermediate states, is $\Gamma/2\pi\approx 100\,\rm{kHz}$. This brings the atom either back to the original ground state $\ket{\downarrow}$ or to auxiliary shelving states $\ket{s}$ external to the spin-1/2 description, such as the $\ket{4s_{1/2},F=1}$ state, with an estimated probability $b=0.18$ (Appendix~\ref{app:excitation_spectrum}). This ultimately leads to all the population accumulating in $|s\rangle$. However, interesting quasi-steady states can be reached for significant periods of time even when $b\neq 0$. 

To probe the open system dynamics we take advantage of the slow loss of population out of the spin-1/2 subspace. Starting with all atoms initially prepared in the ground state, i.e. the fully magnetized state with $m=0$, we suddenly switch on the driving field with fixed values of $\Omega$ and $\Delta$ corresponding to different regions of the non-equilibrium phase diagram [Fig.~\ref{fig:1_regimes}(b)]. Following this quench, the system evolves towards states with $m > 0$ on a timescale set by the inverse decay rate $\Gamma^{-1}$. To probe the (quasi) steady-state established after this short transient period, we let the system evolve for a time $\tau$ that is orders of magnitude longer: between $1.4\,\mathrm{ms}$ and $146.5\,\mathrm{ms}$ (depending on $\Omega$). We then switch off the driving field and using absorption imaging we measure the remaining fraction of ground state atoms $P(\tau)/P(0)$. This is then converted to a rate by assuming exponential decay $R=-\tau^{-1}\ln[P(\tau)/P(0)]$, which we verify is a reasonable assumption for the timescales probed in our experiment (see Appendix~\ref{app:exponential_decay}). In the experimentally relevant situation of $R\ll \Omega,\Gamma$ the rate of decay is a good measure for the (quasi) steady-state magnetization according to $R\approx b\Gamma m$. By adapting $\tau$ for each value of $\Omega$ to limit the maximum lost fraction to $\lesssim 0.5$, it is possible to measure $R$ with a dynamic range of over four orders of magnitude. 
 
To verify that the loss rate is indeed sensitive to the many-body state of the system, we measure its dependence on the driving strength and detuning (Fig.~\ref{fig:2_spectra}). For the smallest $\Omega$ we observe an approximately symmetric Gaussian lineshape with a full width half maximum of $3\,\rm{MHz}$ [Fig.~\ref{fig:2_spectra}(inset)]. This is an order of magnitude broader than the estimated dephasing rate, but is consistent with single-atom master equation calculations including the inhomogeneous light shifts produced by the dipole trap laser (see Appendix~\ref{app:excitation_spectrum}). This agreement is expected, as for small driving amplitudes the density of Rydberg excitations remains small enough that the gas is effectively non-interacting. As $\Omega$ is increased however, the width of the loss resonance grows and becomes noticeably asymmetric towards positive detunings, which is a clear consequence of the strong repulsive interparticle interactions between Rydberg $ns$ states. This is compatible with previous experiments which directly measured the Rydberg excitation fraction and observed asymmetric broadening for short excitation pulses~\cite{Schempp2014, Malossi2014}. Our maximal observed loss rate of $0.15\,\rm{kHz}$ is well below the non-interacting saturated limit $(\Gamma/2\pi)b/2=9\,$kHz which is another indication that interactions strongly influence the loss dynamics.

\section{Non-equilibrium scaling laws}

To exploit the high dynamic range of the loss measurements we now analyse the data for fixed detunings as a function of $\Omega$ as shown in Fig.~\ref{fig:3_scaling}. Generally, the data can be empirically described by multiple power-laws $R\propto \Omega^\alpha$, each spanning one or more orders of magnitude in $R$ and $\Omega$ (shaded lines). The exponents change depending on the detuning and the range of driving field strengths. In the remainder of the paper we show that these power-laws and the boundaries in-between can be used to experimentally distinguish qualitatively different regimes of the driven-dissipative system anticipated from Fig.~\ref{fig:1_regimes}(b).

For detunings far above or below resonance and for weak driving, the data exhibits power law scaling with $\alpha\approx 2$ over most of the measurement range [Fig.~\ref{fig:3_scaling}(a)]. This is consistent with paramagnetic behavior expected from mean-field theory.

Driving the system close to resonance [Fig.~\ref{fig:3_scaling}(b)], when going from small to large $\Omega$ we observe a transition from $\alpha\approx 2$ to a weaker exponent ($\alpha<2$). The best fit exponent for large $\Omega$ has a mean of $\alpha = 1.185$ and standard deviation of $0.025$ for detunings in the range $\pm 1\,\mathrm{MHz}$. The threshold between these two scaling behaviors, determined from a piecewise powerlaw fit, occurs around $\Omega_{\rm{th}}/2\pi \approx 12\,\rm{kHz}$. While the precise value of the scaling exponent differs from the mean-field expectation for the critical regime $\alpha_\mathrm{MF}=2/5$, the change in scaling behavior is similar to the crossover from the dissipation dominated to the critical regime associated to the quantum critical point at $\Delta=\Omega=0$.

For intermediate detunings above resonance, we observe a continuous increase of the loss rate with Rabi frequency which appears to obey a power-law with stronger scaling ($\alpha>2$) compared to all other regimes [Fig.~\ref{fig:3_scaling}(c)]. This is highly suggestive of the effect of an instability driven by fluctuations (in the sense of a continuous phase transition) as opposed to the bistable phase predicted by mean-field theory. To test if the data is indeed well described by a power-law scaling we compare fits to both power-law and exponential growth models. For the data shown in Fig.~\ref{fig:3_scaling}(c) we obtain $\alpha=\nobreak 2.81(9)$ with a reduced $\chi^2=0.63$ for the power-law model which is favored over the exponential model with $\chi^2=11.3$. Power-law scaling with exponents $\alpha=2.8\pm 0.4$ is observed over a wide range of detunings above resonance ($\Delta/2\pi\approx 7 \pm 3\,\rm{MHz}$) as well as a smaller range below resonance ($\Delta/2\pi\approx -3.5 \pm 1.5\,\rm{MHz}$). 

A microscopic mechanism that can explain the stronger scaling above resonance is facilitated excitation. In the presence of an initial seed excitation there is a characteristic distance at which the interaction energy precisely matches the laser detuning, subsequently facilitating further excitations in a runaway process~\cite{Schempp2014,Malossi2014,Urvoy2015,Simonelli2016,Valado2016}. In our data this is very pronounced above resonance, but is also seen slightly below resonance, most likely due to the inhomogeneous light shifts of the optical dipole trap or slight anisotropies of the interaction potential at short interatomic distances. Earlier experimental studies under similar experimental conditions have observed these facilitated excitation processes through super-Poissonian number fluctuations and temporal dynamics~\cite{Schempp2014, Malossi2014,Simonelli2016}. Power law scaling is a newly observed feature of these dynamics.

\begin{figure}[t]
	\centering
	\includegraphics[width = 0.85\columnwidth]{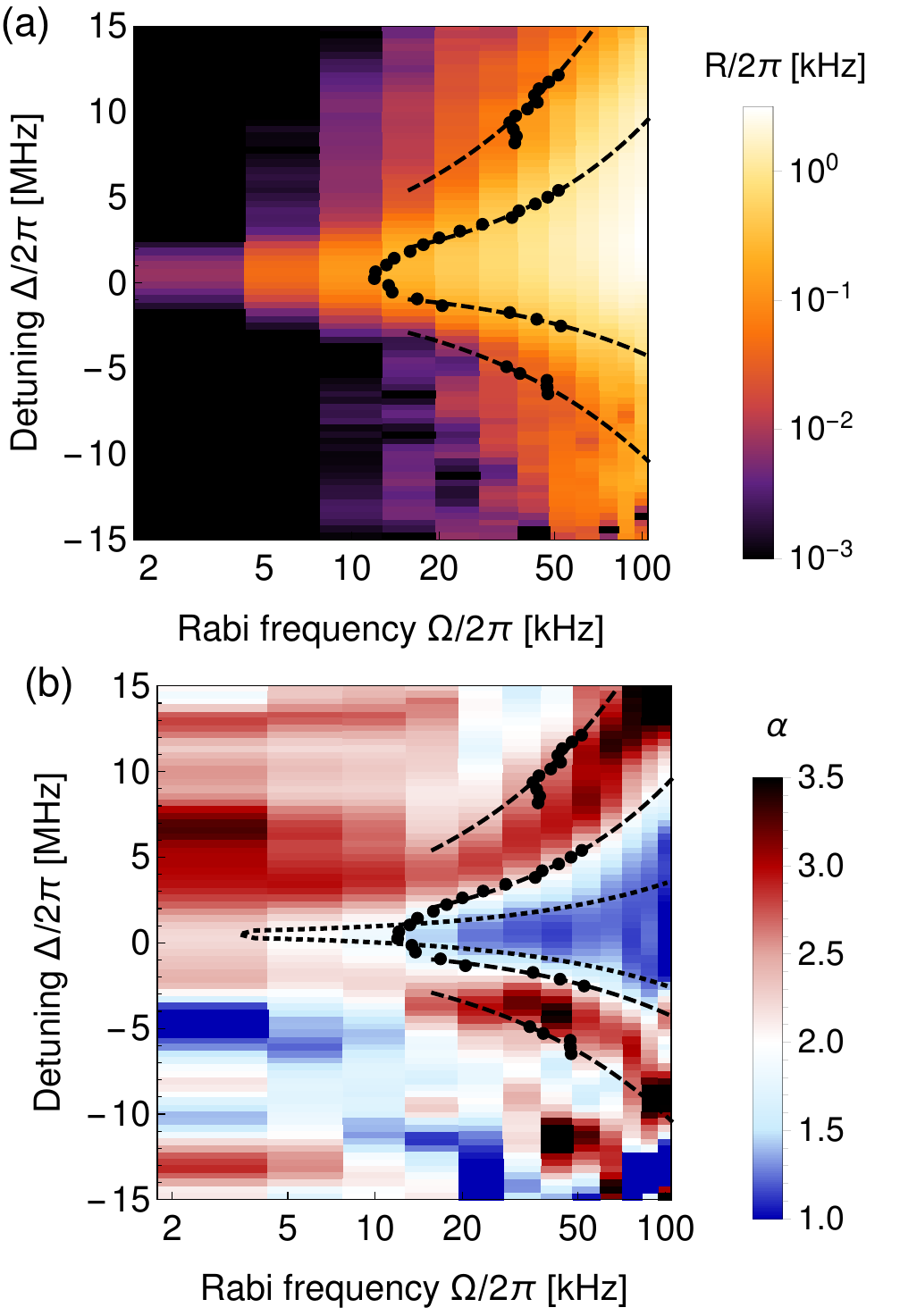}
	\caption{Experimentally measured non-equilibrium phase diagram. (a) As a function of the driving parameters ($\Delta,\Omega$) the loss rate $R$ is fairly smooth with few distinguishing features. (b) In contrast, the locally determined scaling exponent $\alpha$ shows clearly distinguished regimes. The critical regime is identified as the blue region ($\alpha\approx 1.2$), while the unstable regime is red ($\alpha> 2$). The black points and dashed lines show the estimated boundary positions, the dotted line indicates the boundary to the critical regime predicted from mean-field theory corresponding to the highest density region of the cloud (trap center). }
	\label{fig:4_phases}
\end{figure}

For the largest Rabi frequencies reached $\Omega/2\pi \gtrsim 50~$kHz we observe another crossover to weaker scaling ($\alpha < 2$) which is connected to the same critical regime observed on resonance [as can be more clearly seen in Fig.~\ref{fig:4_phases}(b)]. This crossover appears for driving fields and detunings outside the window associated to the critical regime in Fig.~\ref{fig:1_regimes}(b), indicating that runaway facilitation processes tend to drive the system to criticality over a wider parameter range than predicted by mean-field theory [seen by the dotted line in Fig.~\ref{fig:4_phases}(b)].

\section{Coupled rate-equation modelling}

To quantitatively describe the experiment and further elucidate the origin of the different scaling behaviors we go beyond mean-field theory, turning to numerical rate equation (RE) simulations. The basic idea of the RE approach is to describe the excitation dynamics in terms of stochastic jumps between classical spin configurations approximated by single-spin transition probabilities. RE models enable efficient simulation of the steady-state and dynamics of large systems comparable to those realized in experiment and have proven very successful in reproducing the behavior of driven-dissipative Rydberg systems in the presence of dephasing~\cite{Ates2007,Hoening2014,Schempp2014,Gaerttner2014,Valado2016}.

Although atomic motion and the non-conservation of population cannot be fully neglected in our experiments, for simplicity we start with a quasi-static model where each atom is treated with a fixed position and consists of the ground and Rydberg states only. We also include single atom dephasing with rate $\gamma_{\rm{de}}/2\pi\approx 300\,$kHz, compatible with the combined effects of laser linewidth and residual Doppler broadening in our experiment. 
By eliminating off-diagonal elements (coherences) of the density matrix in Eq.~\eqref{eq:lindblad}~\cite{Marcuzzi2014} we obtain the single spin jump rates $\Gamma_{\uparrow,\downarrow}$ that have the form
\begin{equation}\label{eq:rate}
\Gamma_\uparrow^j= \frac{\Omega^2 (\Gamma+\gamma_{\rm{de}})}{(\Gamma+\gamma_{\rm{de}})^2+4(\Delta-V_j)^2}, \quad \Gamma_\downarrow^j=\Gamma_\uparrow^j+\Gamma
\end{equation} 
where $V_j=\sum_{k} V_{jk}n^k$ accounts for van der Waals interactions on atom $j$ depending on the instantaneous configuration of all other spins. We approximate the slow loss out of the two-level subspace by evolving the classical rate equations from the fully magnetised state until the average fraction of Rydberg excitations has converged to its asymptotic value. This is then multiplied by $b\Gamma$ to recover the loss rate $R$. 

To quantitatively describe the data shown in Figs.~\ref{fig:2_spectra},\ref{fig:3_scaling}, we found it necessary to modify the static RE model to include the effect of thermal atomic motion, which enhances the excitation probability for $\Delta>0$ due to the possibility for Landau-Zener transitions. This can be incorporated in the RE model by adding a second term to $\Gamma_\uparrow$ in Eq.~\eqref{eq:rate} accounting for the velocity dependent Landau-Zener transition probability and the Maxwell-Boltzmann velocity distribution as well as the finite lifetime of the Rydberg state (see Appendix~\ref{app:motion}). Without this term, the RE simulations were unable to reproduce stronger scaling observed above resonance within the experimentally accessible range of Rabi frequencies. Finally, we account for the optical trap by averaging the simulated loss rates over the distribution of local detunings and atomic densities assuming Gaussian profiles for the trap laser and atomic cloud (see Appendix~\ref{app:excitation_spectrum}).

Results of the RE simulations are presented alongside the experimental data in Fig.~\ref{fig:3_scaling}. The RE model reproduces the key features of the experimental data over essentially the full range of parameters explored. For example, close to resonance [Fig.~\ref{fig:3_scaling}(b)] the RE model clearly shows the change in scaling around $\Omega_{\rm{th}}$, transitioning from $\alpha=2$ to a weaker scaling close to the experimental value of $\alpha=1.185$. The RE model however does not appear to exhibit quite as sharp a transition as in the experiment.
To investigate further, we also present mean-field and RE simulation results for a homogeneous system focusing on a region close to the center of the trap [dotted line and crosses in Fig.~\ref{fig:3_scaling}(b) respectively]. Here the power-law scaling is even more pronounced with an exponent $\alpha=0.4$, in agreement with the mean-field quantum critical exponent in three dimensions. It may be surprising at first that the RE model, which neglects quantum coherences, can reproduce this scaling behaviour. We attribute this to the fact that we concentrate on a relatively simple observable (the global magnetization) in the long time limit after which any observable effects of coherent dynamics are effectively washed out. Based on this agreement between mean-field theory and the RE simulations as well as between the RE simulations and the experiment, we confirm that the change in scaling occuring around $\Omega_{\rm{th}}$ is a consequence of the transition from dissipative to critical behavior linked to the equilibrium critical point of the Ising-like model. 

Interestingly, the crossover found in experiment and theory occurs for Rabi frequencies significantly below the dissipation rate, which in mean-field theory is set by $\sqrt{\Gamma(\Gamma+\gamma_{\rm{de}})}$. This can be understood as a collective effect arising from the Rydberg blockade, similar to the crossover from weak-coupling to collective strong-coupling regimes of cavity quantum electrodynamics~\cite{Brennecke2007,Colombe2007}. Assuming the crossover occurs at the point where the collectively enhanced driving strength exceeds the total dissipation rate $\sqrt{N}\Omega_{\rm{th}} \approx \sqrt{\Gamma(\Gamma+\gamma_{\rm{de}})}$ we estimate the number of participating atoms to be around $N\approx 300$. This is quite small compared to an independent estimate of the peak number of atoms per Rydberg blockade volume $N\approx(4\pi/3)[2J/(\Gamma+\gamma_{\rm{de}})]^{1/2}=2400$ based on the three-dimensional blockade volume~\cite{Loew2012}. This indicates that the relatively low density wings of the atomic cloud play a dominant role in determining the loss rate. It also shows the importance of the $\sqrt{N}$ enhancement of the atom-light coupling, previously observed for nearly isolated Rydberg superatoms~\cite{Dudin2012,Zeiher2015}, on the non-equilibrium phase structure of the driven-dissipative system. 

The RE model also qualitatively reproduces the stronger scaling ($\alpha>2$) of the loss rate found above resonance as a consequence of the instability (further enhanced by atomic motion). This is seen in the RE simulations in [Fig.~\ref{fig:3_scaling}(c)] as an upwards trend around $\Omega/2\pi\gtrsim 5~\mathrm{kHz}$ for positive detunings. Fitting the RE simulations over a similar range of $\Omega$ and $\Delta$ as for the experimental data yields an exponent of $\alpha=2.40\pm 0.1$, while neglecting trap averaging $\alpha=3.2\pm 0.2$), which spans the range of values found experimentally. The RE model also reproduces the crossover to weaker scaling for large $\Omega$ connected to the critical regime [Fig.~\ref{fig:3_scaling}(c)]. 

\section{Conclusion}\label{sec:conclusion}
We have demonstrated that the rate of population loss and associated scaling laws provide a convenient and robust way to identify vastly different regimes of strongly-interacting open quantum systems. Using the experimentally observed scaling we can even map out the non-equilibrium phase diagram as a function of the control parameters $\Omega$ and $\Delta$, as illustrated in Fig.~\ref{fig:4_phases}. To produce this phase diagram, the scaling exponents are obtained from the slopes of linear fits to $R(\Omega)$ in a moving window on a log-log scale. To locate the boundaries between the regimes we fit connected piecewise linear functions to the log-log-scaled loss rates for each detuning. 
In contrast to the loss rate $R$, which has relatively few distinguishing features [Fig.~\ref{fig:4_phases}(a)], the scaling exponents show distinct regimes corresponding to the critical (blue) and unstable (red) scaling regimes [Fig.~\ref{fig:4_phases}(b)]. The dissipation dominated regime appears as a mostly white region within the detuning interval $|\Delta|/2\pi < 2~\mathrm{MHz}$ and for $\Omega/2\pi\lesssim 12\,\mathrm{kHz}$. Looking at larger positive-to-negative detunings we note a slight trend from red-to-blue is apparent, which might be evidence of a weak interaction effect on the paramagnetic state.

To conclude, scaling laws found in the rate of population decay have made it possible to uncover the non-equilibrium phases of an Ising-like open quantum spin system governed by the competition between driving, dissipation and interactions. We show that the phase structure is extremely rich, exhibiting features which can be attributed to the equilibrium quantum Ising model, i.e. critical scaling in the regime where driving and interactions dominate, but also genuinely new non-equilibrium features, e.g. the collectively enhanced crossover from the dissipation-dominated to the critical regime and an instability towards strongly-correlated states for positive detunings. While the former appears to be captured by mean-field theory which approximates two-point correlations, the latter could only be adequately described using rate equation simulations including many-body correlations. The observed scaling laws also appear to be quite robust, as we find qualitatively similar behavior for a wide range of Rydberg states including $ns$ and $nd$ states which posses different interaction strengths and anisotropies. Thus, we expect they will serve as a powerful tool for identifying universal and non-universal aspects of non-equilibrium quantum systems and as a benchmark for future many-body theories. Future experiments aiming to learn more about the microscopic origins of this scaling behavior, especially in the unstable regime, could reveal the build up of spatial correlations between the spins (e.g. using high resolution imaging techniques for Rydberg atoms~\cite{Schauss2012,Guenter2013}) or look for possible self similar dynamics in the transient evolution~\cite{Gutierrez2015}.

\acknowledgements{We acknowledge discussions with S. Jochim, I. Lesanovsky, G. Lochead, J. Schachenmayer, M. Weidem\"uller and H. Weimer, as well as V. Ivannikov and E. Pavlov for contributions to the experimental setup. This work is supported by the Deutsche Forschungsgemeinschaft under WH141/1-1 and is part of and supported by the DFG Collaborative Research Centre ''SFB 1225 (ISOQUANT)'',  the Heidelberg Center for Quantum Dynamics, the European Union H2020 FET Proactive project RySQ (grant N. 640378) and the `Investissements d'Avenir' programme through the Excellence Initiative of the University of Strasbourg (IdEx). S.H. acknowledges support by the Carl-Zeiss foundation, A.A. acknowledges support by the Heidelberg Graduate School for Fundamental Physics.} 

\appendix

\makeatletter 
\section{Mean-field approximation including two-point correlations}\label{app:mean_field}

Mean-field theory provides a relatively simple way to qualitatively understand the different phases of the driven dissipative system by providing analytic expressions for the magnetization and corresponding scaling laws. The mean-field equations can be obtained by assuming each spin interacts with the average field produced by all other spins. However, to account for the strong correlations which arise due to the Rydberg blockade effect in a self-consistent way, we include a distance dependent cutoff to the interaction term. This model has previously been shown to reproduce the quantum critical behavior associated with the antiferromagnetic Ising model in the absence of dissipation~\cite{Weimer2008,Loew2009, DeSalvo2016}. Here we extend this description to open system dynamics, by applying the self-consistent mean-field approximation to the full master equation [Eq.~\eqref{eq:lindblad}]. In the following we briefly describe the derivation of this model and discuss its solution in different limits, focusing on the quasi-steady state magnetization within the two-level subspace. 

We start by applying the mean-field approximation to the interaction term of the Hamiltonian:
\begin{align}
\frac{1}{2}	\sum_{j,k \neq j} V_{jk} n^j n^k \rightarrow \sum_{j} V_j n^j, 
\end{align}
where 
\begin{align}
V_j= \sum_{k\neq j} V_{jk} m g_2\big(|\mathbf{r_j}-\mathbf{r_k}|\big).
\end{align}
This describes the effective interaction of a single spin with all other spins weighted by the local density of Rydberg excitations. Following Weimer {\emph{et al.}} \cite{Weimer2008}, we approximate the two-point correlation function by the  Heaviside step function $g_2(\mathbf{|r|}) = \Theta(\mathbf{|r|}-r_\mathrm{c})$, where $r_\mathrm{c}$ is a characteristic distance for the correlations. Next we replace the discrete sum by an integral with homogeneous density $\rho_0$,
\begin{eqnarray}
V_j\approx& \int_{r_\mathrm{c}}^\infty m \rho_0V_{jk}  \;\mathrm{d}^d{\mathbf{r_k}},
\end{eqnarray}
which is a good approximation in the experimentally relevant situation that the the mean interparticle distance is much less than $r_\mathrm{c}$. Subsituting $V_{jk}=C_p/|\mathbf{r_j}-\mathbf{r_k}|^p$ (where $p=6$ for van der Waals interactions) and using the spherical symmetry allows one to carry out the integration explicitly. Furthermore, we self-consistently set $r_\mathrm{c}$ by requiring that there is on average only one excitation within the correlation distance due to Rydberg blockade, $\int\rho_0 m \big(1-g_2(\mathbf{r})\big) \;\mathrm{d}^d \mathbf{r} \equiv 1$, yielding:
\begin{align}
V_j =
\frac{C_p d\,{(m \rho_0 \mathcal{V}_d)}^{p/d}}{p-d} = J c\, {m}^{p/d}, 
\end{align}
where we have introduced the volume of the $d$-dimensional unit sphere $\mathcal{V}_d$, $J = C_p {\rho_0}^{p/d}$ parametrises the interaction strength and $c= {\mathcal{V}_d}^{p/d} d/(p-d)$ is a dimensionless constant. Thus the mean-field Hamiltonian, using $n=(\sigma_z+1)/2$ and dropping a constant energy offset, is 
\begin{align}
H = \frac{\Omega}{2}\sigma_x - \frac{1}{2}\left(\Delta- Jc\, {m}^{p/d}\right) \sigma_z, 
\end{align}
which can be readily inserted into the master equation~\eqref{eq:lindblad} for the single-spin density matrix $\rho$. The mean-field master equation is solved self-consistently for the Rydberg population $m = \mathrm{Tr}[\rho \,n]$, whose solution can be expressed implicitly for small Rydberg populations as 
\begin{align}
\Omega^2 \approx  \biggl{(}4 \bigl(\Delta-Jc\, {m}^{p/d} \bigr)^2 + (\Gamma+\gamma_{\rm{de}})^2\biggr{)} \frac{m\Gamma}{\Gamma+\gamma_{\rm{de}}} . 
\end{align}
The numerical solution to this equation (for $\Delta=0$) is depicted as a solid line in Fig.~\ref{fig:3_scaling}(b). 

Depending on the dominant energy scales we find the following regimes as presented in the main text: \\ 
\emph{Paramagnetic regime} ($|\Delta|\gg Jc\,{m}^{p/d};\;\Gamma+\gamma_{\rm{de}}$): 
\begin{align}
m = \left( \frac{\Omega}{2\Delta} \right)^2 \frac{\Gamma+\gamma_{\rm{de}}}{\Gamma}
\end{align}
\emph{Dissipation-dominated regime} ($\Delta \approx 0;\; \Gamma+\gamma_{\rm{de}} \gg Jc\,{m}^{p/d}$): 
\begin{align}\label{eq:mdiss}
m = \frac{\Omega^2}{\Gamma (\Gamma+\gamma_{\rm{de}})}
\end{align}
\emph{Critical regime} ($\Delta \approx 0;\; Jc\,{m}^{p/d}\gg \Gamma+\gamma_{\rm{de}}$): 
\begin{align}\label{eq:mcrit}
m = \left( \frac{\Omega}{2Jc} \sqrt{\frac{\Gamma+\gamma_{\rm{de}}}{\Gamma}} \right)^{1/\delta} \quad \mathrm{with} \quad \delta = \frac{p}{d}+\frac{1}{2}
\end{align}
Thus the mean-field magnetization scales as $\Omega^{1/\delta}$ in the critical regime, which is the equilibrium scaling exponent. \\
\emph{Unstable regime} ($|\Delta-Jc\,{m}^{p/d}|\; \gg \Gamma+\gamma_{\rm{de}}$): In the regime where dissipation is small yet neither the interactions nor the detuning clearly dominate, the magnetization takes on multiple solutions with low and high Rydberg fraction, according to:
\begin{align}
\Omega = 2\sqrt{m}\left| \Delta-Jc\,{m}^{p/d}  \right| \sqrt{\frac{\Gamma}{\Gamma+\gamma_{\rm{de}}}}
\end{align}

From the above formulas, expressions for the boundaries between the regimes can also be derived. As an example, we will discuss the threshold between driving and dissipation dominated regimes found by equating Eq.~\eqref{eq:mdiss} and Eq.~\eqref{eq:mcrit}: 
\begin{align}
\Omega_\mathrm{th} = \sqrt{\Gamma(\Gamma+\gamma_{\rm{de}})} \left(\frac{\Gamma+\gamma_{\rm{de}}}{2 J c}\right)^{d/2p}  
\end{align}
The last factor in the equation above can be expressed in terms of the $d$-dimensional blockade volume $\mathcal{V}_\mathrm{bl}=\nobreak\mathcal{V}_d (\frac{2C_p}{\Gamma+\gamma_{\rm{de}}})^{d/p}$ containing $N = \rho_0 \mathcal{V}_\mathrm{bl}$ atoms~\cite{Loew2012}. Thus we find for the crossover position:
\begin{align}
\sqrt{N} \Omega_\mathrm{th} = \sqrt{\frac{\Gamma(\Gamma+\gamma_{\rm{de}})}{\left(p/d-1\right)^{d/p}}} 
\end{align}

Taking the experimental parameters corresponding to the center of the atomic cloud: $N=2400$, $\Gamma/2\pi=100\,$kHz, $\gamma_\mathrm{de}/2\pi=300\,$kHz, we obtain $\Omega_\mathrm{th}/2\pi=4\,$kHz, compatible with the change in scaling observed for the homogeneous theory results shown in Fig.~\ref{fig:3_scaling}(b).

\section{Experimental methods} \label{app:experimental_methods}

To couple the ground state $\ket{\downarrow}=\ket{4s_{1/2},F=2,m_F=2}$ and the Rydberg state $\ket{\uparrow}=\ket{66s}$ we use a coherent two-photon excitation scheme. This utilises a weak `probe' laser at $767\,\mathrm{nm}$ wavelength for the lower $\ket{\downarrow}\leftrightarrow \ket{e} = \ket{4p_{3/2}}$ transition and a strong `coupling' laser at $456\,\mathrm{nm}$ wavelength for the upper $\ket{e} \leftrightarrow \ket{\uparrow}$ transition. The atomic cloud has $e^{-1/2}$ radii $\{\sigma_r,\sigma_z\} = \{7\,\rm{\mu m},220\,\rm{\mu m}\}$. The probe laser is aligned perpendicularly to the long axis with a waist of $\approx 10\,\rm{mm}$. It homogeneously illuminates the cloud with tunable Rabi frequencies $\Omega_p/2\pi\lesssim 1.2\,\rm{MHz}$ and detunings from the $\ket{e}$ state in the range $\Delta_p/2\pi=80\pm 20\,\rm{MHz}$. The coupling laser is derived from a frequency doubled Ti:Sa laser with a total power of 1.15~W focused to a waist of approximately $30\,\rm{\mu m}$ and is aligned collinearly with the long axis of the atom cloud for maximum homogeneity. In our measurements we keep the coupling Rabi frequency and the coupling laser detuning fixed to the values $\Omega_c/2\pi\approx 20\,\rm{MHz}$, $\Delta_c/2\pi=-77\,\rm{MHz}$ and tune $\Omega_p$ and $\Delta_p$. The combined linewidth of both lasers is $\lesssim 200\,\rm{kHz}$. 

The three-level system can be reduced to an effective two-level system (with driving parameters $\Omega, \, \Delta$) in the limit that the population in $\ket{e}$ is negligibly small, e.g. in the limit $|\Delta_p| \approx|\Delta_c|\gg \Omega_c \gg \Omega_p$. For $\Delta_p \approx -\Delta_c$ we can describe the coupling from $\ket{\downarrow}\leftrightarrow \ket{\uparrow}$ by an effective two-photon Rabi frequency $\Omega\approx \varepsilon\Omega_p $, where $\varepsilon=\Omega_c /(2|\Delta_c|) \ll 1$, and an effective detuning $\Delta=\Delta_p+\Delta_c+\Delta_\mathrm{ODT}$, where $\Delta_\mathrm{ODT}$ accounts for the additional light shifts from the $1064\,\mathrm{nm}$ optical dipole trap laser. The position of maximum loss is additionally shifted by $0.5\,\mathrm{MHz}$ due to the light shift produced by the coupling laser and averaging over the inhomogeneous optical dipole potential. The effective excited state decay rate $\Gamma/2\pi\approx (\Gamma_r+\varepsilon^2\Gamma_e)/2\pi \approx 100\,\rm{kHz}$ is a combination of the bare Rydberg state decay $\Gamma_r/2\pi\approx 1.2\,\rm{kHz}$ (including blackbody transitions) and the residual intermediate state admixture which spontaneously decays with rate $\Gamma_e/2\pi=6.03\,\mathrm{MHz}$. Additional loss processes for $ns$-Rydberg states, e.g., photoionization or penning ionization are estimated to be below $1\,$kHz \cite{Potvliege2016} and can be neglected.

Our measurements are performed by varying the probe laser intensity over the range from $30\,\rm{nW/cm}^2$ to $30\,\rm{\mu W/cm}^2$, which corresponds to $\Omega/2\pi$ from $3\,\rm{kHz}$ to $100\,\rm{kHz}$. The atom-light interaction time $\tau$ is set by pulsing an acousto-optical modulator for the probe laser while the coupling laser is kept on continuously.

\section{Reproducing the low intensity excitation spectrum}\label{app:excitation_spectrum}

Before performing RE simulations for the full many-body system we calibrate the model against the low intensity loss rate data shown in Fig.~\ref{fig:2_spectra}(inset). For this we take the analytic solution for the single-atom steady state Rydberg population multiplied by a factor $b\Gamma$. However, the experimental spectrum is significantly broadened because of inhomogeneous level shifts mainly originating from the optical dipole trap laser. Therefore, we convolve the simulated spectra by the energy distribution of atoms in the trap, parameterized by the ratio of the cloud radius to the waist of the optical dipole trap $\sigma/w$. We fix the dephasing rate (accounting for the combined effect of laser linewidth and motional dephasing) to $\gamma_{\rm{de}}/2\pi=300$~kHz and adjust the unknown parameters $b=0.18$, $\Delta_\mathrm{ODT}/2\pi=4.0\,$MHz and $\sigma/w=0.30$ to obtain best agreement with the data within the known experimental constraints. Careful inspection of the experimental loss rate data additionally reveals an unexpected broad pedestal centered around $\Delta/2\pi\approx 2.1~$MHz with a standard deviation of $8.7~$MHz but with an amplitude one order of magnitude smaller than the peak loss rate. We include this pedestal into the RE model as an additional single atom excitation process.

\section{Exponential population decay}\label{app:exponential_decay}

\begin{figure}[t]
 \centering
 \vspace{10pt}
	\includegraphics[width = .8\columnwidth]{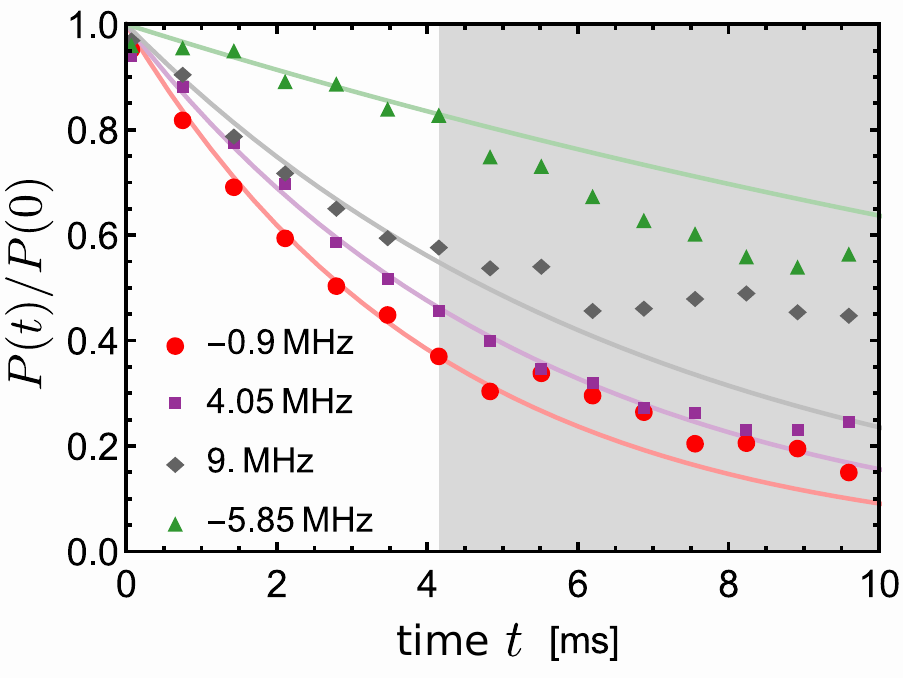}
	\caption{Time resolved measurements of the fraction of atoms remaining in the ground state after an excitation pulse of varying duration. Data for a fixed Rabi frequency of $\Omega/2\pi=22~$kHz and four different detunings are shown. The boundary between white and grey backgrounds marks the waiting time $\tau$ used in all experiments for this value of $\Omega$.}
	\label{fig:AtomLoss_dynamics}\vspace{-10pt}
\end{figure}

Each datapoint for the loss rate is inferred from only two population measurements assuming exponential population decay. Here we show this is a good assumption when the maximum lost fraction is $P(\tau)/P(0)\lesssim 0.5$. In Fig.~\ref{fig:AtomLoss_dynamics} we present additional time resolved data of the population loss for $\Omega/2\pi=22~$kHz and four different detunings spanning the paramagnetic and unstable regimes as well as the dissipative-critical crossover. According to our measurement protocol, the waiting time $\tau$ used to extract $R$ for this value of $\Omega$ is $4.1$~ms (boundary between white and grey shaded regions). The data in the region $t<\tau$ is very well described by exponential decay curves, confirming that the formula $R=-\tau^{-1}\ln[P(\tau)/P(0)]$ can be used as a good estimator for the loss rate. For different values of $\Omega$ we find similar exponential behavior. For times longer than $\tau$ we note some deviations from the pure exponential decay, but this is of no consequence for the results presented in this paper.

\section{Motional enhancement of the off-resonant excitation probability}\label{app:motion}

For the long times relevant for our experiments, atomic motion can play a significant role in the excitation dynamics, especially for $\Delta>0$. The dominant effect is that, compared to the case of static disorder, motion enables a greater fraction of atoms to meet the resonance condition at the faciliation distance $r_{\rm{fac}}=(C_6/\Delta)^{1/6}$ and to undergo Landau-Zener transitions.

\begin{figure}[!t]
	\centering
	\includegraphics[width = 0.85\columnwidth]{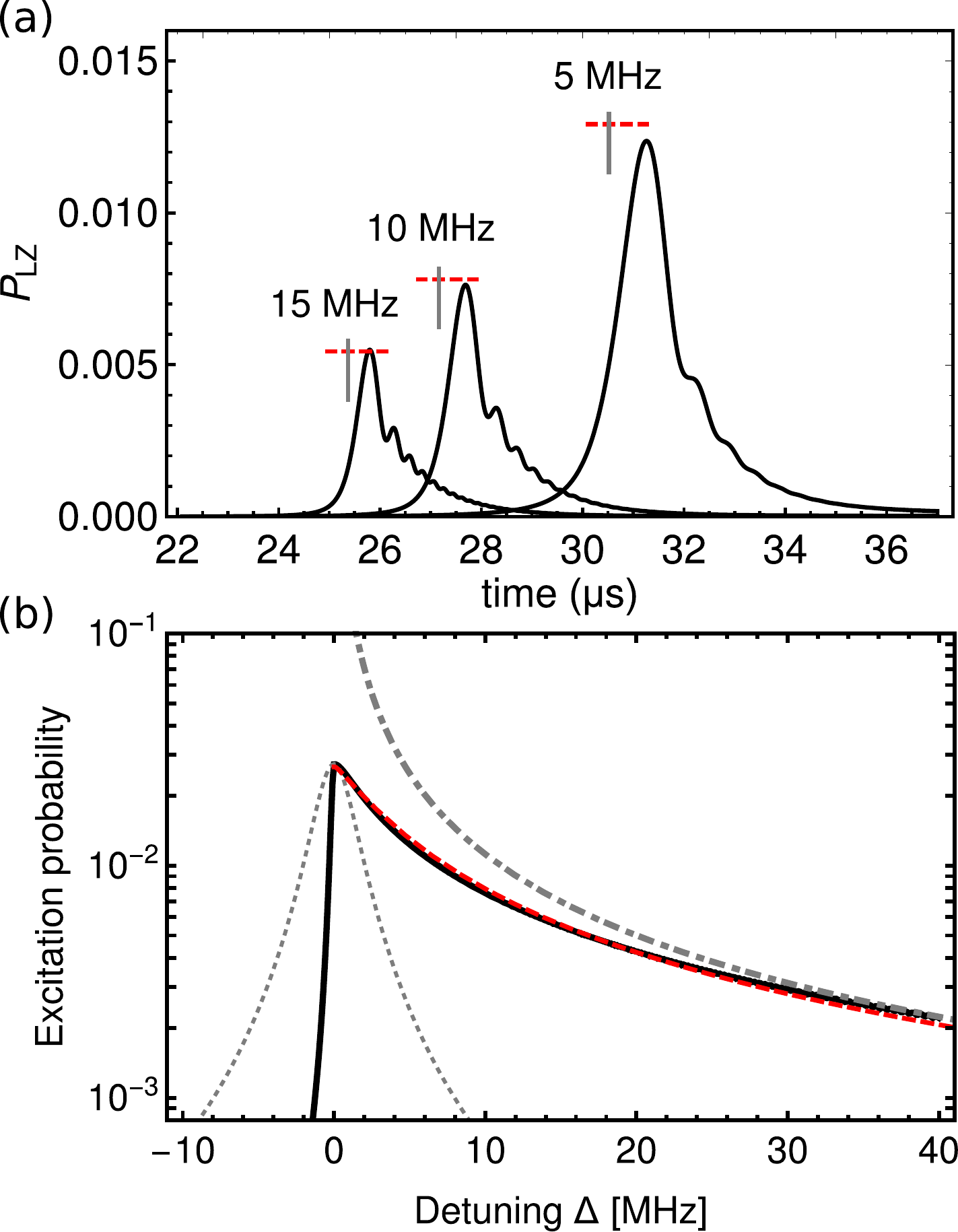}
	\caption{Landau-Zener transition probability for an atom moving away from a Rydberg excitation with a velocity of $0.2\,\mathrm{m/s}$. (a) transient excitation dynamics for three different detunings. The dashed red horizontal lines show the analytic approximation (see text) and the vertical gray lines mark the time at which the atom reaches the facilitation shell for each detuning. (b) Peak excitation probability as a function of detuning. The solid black line shows the full numerical simulations, the red dashed line is the modified Landau Zener theory including dissipation (Eq.~\eqref{eq:landauzener}), the dash-dotted line is the usual Landau-Zener result neglecting dissipation and the dotted line is the steady state Rydberg fraction assuming static atoms.}
	\label{fig:landauzener}
\end{figure}

To account for this motion enhanced excitation probability in the RE simulations we derive an expression for the probability to excite an atom from its ground state while it moves with velocity $v$ relative to a pre-excited Rydberg atom. Their separation $r$ is treated as a classical parameter, while the excitation dynamics of the ground-state atom is described by the quantum master equation (Eq.~\ref{eq:lindblad}), which includes decay and dephasing terms and an effective position dependent detuning $\Delta-V(r)$. Figure~\ref{fig:landauzener}(a) shows examples of the time-dependent excited state probability assuming the two atoms start at the same initial position ($r=0$) and move apart as a function of time, for parameters similar to the experiment and for three different detunings. After the atom crosses the facilitation distance (indicated by gray vertical lines in Fig.~\ref{fig:landauzener}(a) for each detuning), the excitation probability peaks and undergoes oscillatory dynamics which are damped due to spontaneous decay and dephasing. The peak excitation probability is much larger than the off-resonant (static) probability and, depending on the relative velocity of the atoms, can be a sizable fraction of the maximum on-resonant probability [Fig.~\ref{fig:landauzener}(b)].

For the experimental parameters assumed, the typical crossing time is comparable to the dissipation rate making an analytic treatment difficult. However, by inspecting the numerical simulations we found that the peak excitation probability $P_{\rm{LZ}}$ is well described by a heuristic model which incorporates the usual Landau-Zener transition probability $P_{\rm{LZ}}^0$ with a cut-off given by the static resonant excitation probability $f_R$, according to:
\begin{equation}\label{eq:landauzener}
P_\mathrm{LZ}=\left(\frac{1}{f_R}+\frac{1}{P_{\rm{LZ}}^0}\right)^{-1}
\end{equation}
where
\begin{equation}
P_{\rm{LZ}}^0=1-\mathrm{e}^{-2\pi (\Omega/2)^2/|\dot\delta|},\quad f_R=\frac{\Gamma_\uparrow}{\Gamma_\uparrow+\Gamma_\downarrow}\biggl |_{V=-\Delta}
\end{equation}
and $\dot\delta=v(dV/dr)=-6 v\Delta^{7/6}/C_6^{1/6}$ is the slew rate of the Landau-Zener energy level crossing evaluated at the facilitation distance (assuming van der Waals interactions). Fig.~\ref{fig:landauzener}(b) shows a comparison of the peak excitation probability for the modified Landau-Zener probability $P_\mathrm{LZ}$ (dashed red line) and the full numerical simulation of the time-dependent master equation (solid black line) alongside the bare excitation probability without motional enhancement (dotted gray line) and the usual Landau-Zener result without dissipation (dash-dotted gray line). For $\Delta/2\pi \gtrsim 5~$MHz the motion-enhanced excitation probability is more than an order of magnitude larger than the excitation probability without motion.

So far, this treatment does not include the probability that a given atom actually undergoes a Landau-Zener crossing, which depends on it's initial position and velocity as well as the lifetime of the Rydberg state. To account for this we assume a Maxwell-Boltzmann velocity distribution characterised by a thermal velocity $v_\mathrm{th}=(2 k_B T/m)^{1/2}$. For the experimental parameters, the distance the atoms move during the excited state lifetime is small compared to the facilitation distance. Therefore it is sufficient to use a one-dimensional model in which the facilitation shell is treated as a planar boundary at $x=x_\mathrm{fac}$. After integration we find that the probability for a given atom with initial position $x_j$ to cross the boundary within the time $t$ is:

\begin{equation}
\mathcal{P}_{\mathrm{cross}}=\frac{1}{2}\mathrm{erfc}\left( \frac{|x_\mathrm{fac}-x_j|}{t\,v_\mathrm{th}}\right ).
\end{equation}
To account for the possibility that the Rydberg atom decays before the boundary is reached, we time-integrate the crossing probability weighted by an exponential decay

\begin{eqnarray}\label{eq:Pcross}
P_\mathrm{cross}&=&\frac{\int_{0}^\infty ~\mathcal{P}_\mathrm{cross}\mathrm{e}^{-2\pi\Gamma t} dt}{\int_{0}^\infty \mathrm{e}^{-2\pi\Gamma t} dt}\nonumber\\
&\approx& \frac{1}{\sqrt{3}}\exp\left(-3\xi^{2/3}\right)
\end{eqnarray}
where in the last step we use an approximation to the MeijerG special function for $\xi=\pi\Gamma |x_\mathrm{fac}-x_j|/v_\mathrm{th} > 0$ which is accurate within 14\%.

To incorporate this model into the RE simulations we modify Eq.~\eqref{eq:rate} according to $\Gamma_\uparrow\rightarrow \Gamma_\uparrow + \Gamma P_\mathrm{LZ}P_\mathrm{cross}$, substituting $x_\mathrm{fac}\approx(C_6/\Delta)^{1/6}$ and $x_j\approx(C_6/V_j)^{1/6}$. Generally, this prescription underestimates the effect of motion as it doesnt include forces between the atoms or the possibility that there are multiple Rydberg excitations in the viscinity of a given atom. For the results shown in Fig.~3, we find best agreement with the data assuming $P_\mathrm{cross}$ is four times larger than the expression given in Eq.~\eqref{eq:Pcross}.

\bibliography{dressed_scaling}

\end{document}